\documentclass[nofootinbib,twocolumn,aps,pre,superscriptaddress,citeautoscript,floatfix, groupedaddress]{revtex4-1}

\usepackage{graphicx}
\usepackage{amsmath,amssymb}
\usepackage[dvipsnames]{xcolor}
\usepackage{float}

\begin{document}
%%%%%%%%%
\title{Brownian motion of a charged colloid in restricted confinement}
\author{Yael Avni$^{1,2}$}
\author{Shigeyuki Komura$^{2}$}
\author{David Andelman$^{1}$}
\affiliation{${}^{1}$Raymond and Beverly Sackler School of Physics and Astronomy, Tel Aviv University, Ramat Aviv 69978, Tel Aviv, Israel}
\affiliation{${}^{2}$Department of Chemistry, Graduate School of Science, Tokyo Metropolitan University, Tokyo 192-0397, Japan}

%%%%%%%%%%%%%%%%%%%%%%%%%%
%%%%%%%%
\begin{abstract}
We study the Brownian motion of a charged colloid, confined between two charged walls, for small separation between the colloid and the walls. The system is embedded in an ionic solution. The combined effect of electrostatic repulsion and reduced diffusion due to hydrodynamic forces results in a specific motion in the direction perpendicular to the confining walls. The apparent diffusion coefficient at short times as well as the diffusion characteristic time are shown to follow a sigmoid curve as function of a dimensionless parameter. This parameter depends on the electrostatic properties and can be controlled by tuning the solution ionic strength.
At low ionic strength, the colloid moves faster and is localized, while at high ionic strength it moves slower and explores a wider region between the walls, resulting in a larger diffusion characteristic time.
\end{abstract}
%%%%%%%
\maketitle
%%%%%%%%
\section{Introduction}
%%%%%%%%%%

Understanding the Brownian motion of colloids under confinement has been a great challenge in recent decades~\cite{Burada2009}. Such motion is present in the segregation and transport of particles though the bio-cellular membranes~\cite{Phillips2012}, in microfluidic devices~\cite{Squires2005}, and in particle trapping and tracking technologies~\cite{Krishnan2010,Ruggeri2017}.

A simple example of confinement is observed when a colloid is placed in a confined space bounded by rigid surfaces. In this case, the colloid motion differs from its free medium one (no boundaries) due to hydrodynamic forces between the colloid and the confining surfaces. At low Reynolds numbers, this effect reduces to a change in the drag coefficient, which in turn leads to a position-dependent diffusion coefficient (PDDC) of the colloid. For simple geometries such as a spherical colloid near a single planar wall or in between two flat walls, the PDDC was calculated by
Brenner {\it et\ al.} ~\cite{Brenner1961,Goldman1967} and
Gratos {\it et\ al.}~\cite{Ganatos1980Part1,Ganatos1980Part2}.
These results were verified experimentally for micro- or nano-sized particles, by different techniques of light scattering~\cite{Lan1986,Lobry1996,Hosoda1998,Bevan2000}, video microscopy~\cite{Lancon2002}, optical tweezers~\cite{Dufresne2001} and total internal reflection velocimetry~\cite{Huang2007}.

In polar solvents, the colloid and the confining surfaces often carry an electric charge, and the interaction between them confines the colloid even further to the vicinity
of some potential minimum~\cite{Kepler1994}. This results in an intricate motion which is affected both by the PDDC and the confining potential.

Consequently, the colloid motion was examined experimentally in the lateral direction~\cite{Eichmann2008,Fringes2018}. However, its motion in the direction perpendicular to the confining surfaces, where the coupling between the PDDC and the interaction potential is pronounced, was only examined when the diffusion coefficient does not change throughout the colloid motion~\cite{Feitosa1991,Frej1993}. Although this simplification is valid at times, a rigorous description of the motion in which the PDDC and interaction potential are fully coupled to one another is still missing. Moreover, in previous studies, the electrostatic interaction was approximated by the simplified form, $U\sim\exp{(-d/\lambda_{\rm D})}$, where $\lambda_{\rm D}$ is the Debye screening length and $d$ is the separation distance between the surface of the colloid and the wall. While this is valid in the large separation limit $d\gg \lambda_{\rm D}$, the interesting regime where $d$ is of the order of $\lambda_{\rm D}$ has yet to be explored.

In this paper, we study the motion of a spherical charged colloid, of radius $a$, confined between two charged walls, and focus on the motion in the direction perpendicular to the walls. We consider the large sphere limit, i.e., $a\gg\lambda_{\rm D},d$, for which the calculation greatly simplifies and analytical results are obtained. We make an additional simplification by approximating the interaction potential to be a harmonic one around the equilibrium position.
We obtain a dynamical equation that depends on a single dimensionless parameter, $\alpha$, that quantifies the interplay between the walls and the interaction potential.

In the small or large $\alpha$ limits, the calculated probability distribution function agrees with previous results, while for $\alpha\sim1$, the motion deviates substantially from these two known cases. The difference is quantified by the mean square displacement (MSD) behavior as a function of time, where an analytic expression for the short-time behavior is derived, while the long-time behavior is studied numerically. We also show the dependence of $\alpha$ on the electrostatic properties of the system. In particular, we demonstrate how it can be tuned by changing the salt concentration, causing the colloid motion to crossover between the PCCD dominated motion, which is slow and explores a wide region, and the electrostatically dominated one, which is faster and more localized.

The outline of the paper is as follows. In Sec.~\ref{Equilibrium}, we calculate the interaction potential. In Sec.~\ref{Dynamics}, we calculate the PDDC and derive the dynamical equation of the colloid motion. In Sec.~\ref{Results}, we analyze the MSD
of the motion for different values of the parameter $\alpha$, and show the dependence of $\alpha$ on the electrostatic properties of the system. In Sec.~\ref{Discussion}, we conclude with several general observations and comments on the validity of our results.

%%%%%%%%
\section{Electrostatic poterntial} \label{Equilibrium}
%%%%%%%%%%

\begin{figure}
\includegraphics[width = 0.85 \columnwidth,draft=false]{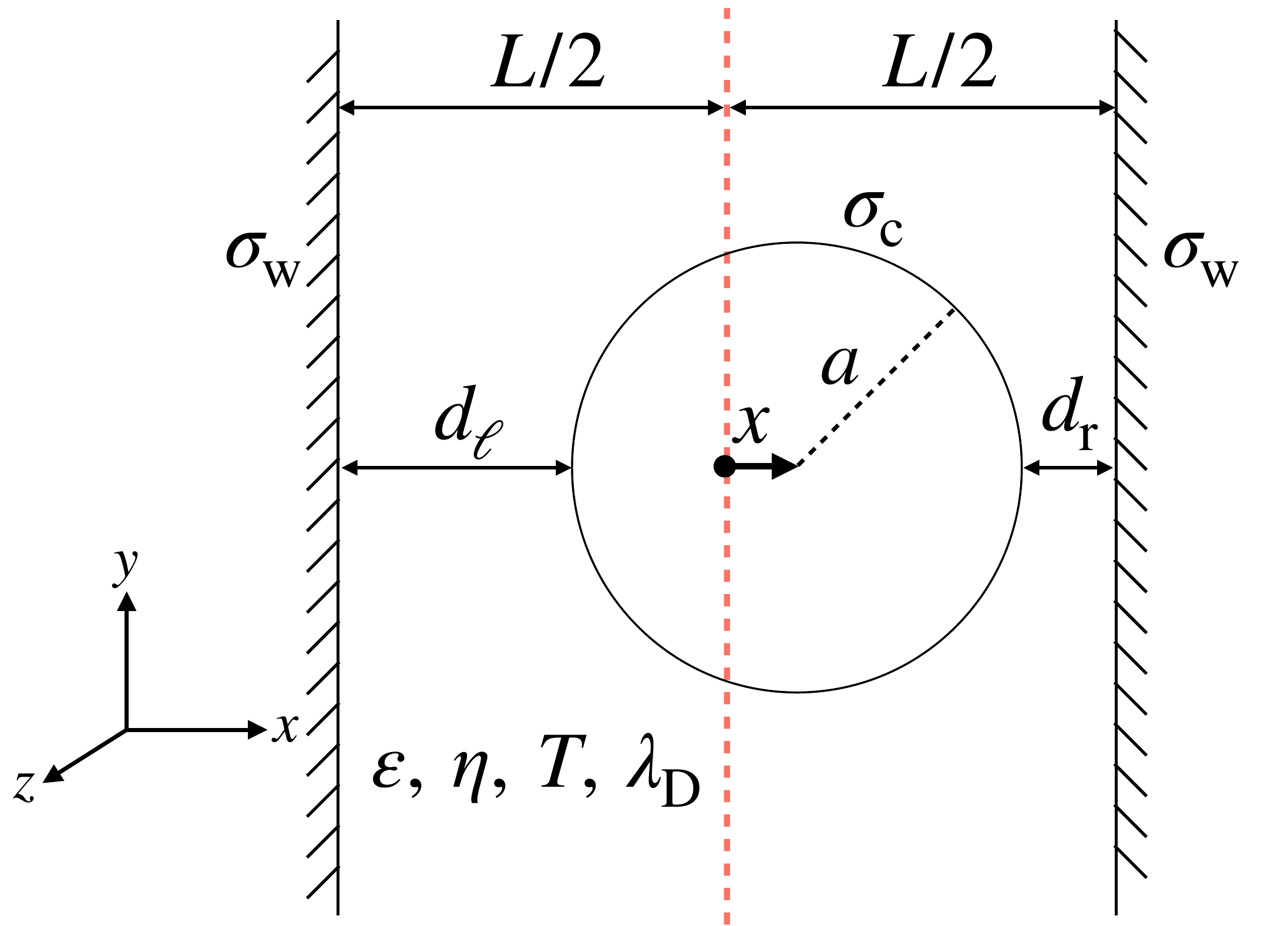} %0.4
\caption{\textsf{Schematic drawing of the system. A colloid of surface charge density $\sigma_{\rm c}$ and radius $a$ is positioned in between two walls with distance $L$ between them, each having a surface charge density $\sigma_{\rm w}$. Both the colloid and walls are embedded in an ionic solution with dielectric constant $\varepsilon$, viscosity $\eta$, temperature $T$ and screening length $\lambda_{\rm D}$.}}
\label{Fig1}
\end{figure}

We consider a spherical colloid of radius $a$ and charge $Q$, embedded in a dilute ionic solution. The charge of the colloid is assumed to be distributed homogeneously on its surface, with surface charge density
$\sigma_{\rm c}=Q/(4\pi a^2)$, where $\sigma_{\rm c}$ can be either positive or negative.
The ionic solution is characterized by a dimensionless dielectric constant $\varepsilon$, viscosity $\eta$, temperature $T$ and bulk concentration of monovalent salt $n$. The colloid is confined between two charged walls, as shown in Fig.~\ref{Fig1}. The distance between the walls is $L$, and the position of the center of the colloid relative to the mid-plane between the two walls is $x$. The distance between the colloid and the left (right) wall is denoted by $d_{\ell}$ ($d_{\rm r}$), such that $d_{\ell} = L/2-a+x$ and  $d_{\rm r} = L/2-a-x$. We define $d_0=L/2-a$ as the distance between
the colloid and the walls when the colloid is placed on the mid-plane ($x=0$).
We also assume that the walls have a fixed charge density, $\sigma_{\rm w}$, which can be either positive or negative.
%%%%%%%%%%%%%%%%%%%%%%%%%%%%%%%%%%%%%%%%%%%
%Fig1
%%%%%%%%%%%%%%%%%%%%%%%%%%%%%%%%%%%%%%%

The electrostatic interaction between a spherical colloid and a flat wall, embedded in an ionic solution, is described
by the Poisson-Boltzmann (PB) theory~\cite{Israelachvili1992,Tomer2020}. Assuming that the colloid and wall are not highly charged
[see the exact condition in Appendix~\ref{appendixA} after Eq.~(\ref{elec_potential})],
the theory can be linearized and reduces to the Debye-H\"uckel (DH) theory. In the DH theory, the electrostatic potential, $\psi$, at each point in space ${\bf r} = (x,y,z)$ is given by the linearized equation
\begin{equation} \label{DH}
(\nabla^2-\kappa_{\rm D} ^2) \psi({\bf r})=0,
\end{equation}
where $\kappa_{\rm D}$ is the inverse screening length, ${\kappa_{{\rm D}}=\lambda_{\rm D}^{-1}=[2e^{2}n/(\varepsilon_{0}\varepsilon k_{{\rm B}}T})]^{1/2}$, $\lambda_{\rm D}$ is the Debye screening length, $e$ is the elementary charge, $\varepsilon_{0}$ is the vacuum permittivity and $k_{\rm B}$ is the Boltzmann constant. The equation is solved together with the boundary conditions of fixed charge on the colloid surface and on the walls.

The DH equation does not have an analytic solution for the geometry considered here. However, by assuming that $a$ is the largest length scale in the system, $a \gg d_{\ell}, d_{\rm r}, \lambda_{\rm D}$ (the large sphere limit), Eq.~(\ref{DH}) can be solved by using the Derjaguin approximation. Note that the large sphere limit combines two different and independent physical limits: $a \gg d_{\ell}, d_{\rm r}$ (narrow confinement) and $a \gg \lambda_{\rm D}$ (thin double layer).

In Appendix~\ref{appendixA}, we calculate the electrostatic potential $\psi$ in the large sphere limit. In addition, the Appendix contains the detailed derivation of the interaction potential, $U(x)$, between the colloid and the charged walls, in the same limit. For convenience, we repeat here only the final expression relating $U(x)$ and the osmotic pressure, $\Pi$
\begin{equation} \label{eq2}
\begin{split}
U(x)=&-2\pi a \bigg(\int\limits^{d_{0}+x}{\rm d}h\int\limits _{h}^{\infty}{\rm d}l\,\Pi\left(l\right)\\
&+\int\limits^{d_{0}-x}{\rm d}h\int\limits _{h}^{\infty}{\rm d}l\,\Pi\left(l\right)\bigg),
\end{split}
\end{equation}
where the osmotic pressure, $\Pi$~\cite{Tomer2020}, is related to $\psi$ by
\begin{align} \label{Pressure2}
\begin{split}
\Pi=\frac{\varepsilon\varepsilon_0}{2}\left(-\psi'^{2}+\kappa_{\rm D}^{2}\psi^{2}\right),
\end{split}
\end{align}
and $\psi'$ is the first derivative along the $x$ direction (note that $U(x)$ is defined up to a constant).

Equations~(2) and (3) (see details in Appendix~\ref{appendixA}) yield the following interaction potential,
\begin{align} \label{Total_potential}
\begin{split}
U (x) & =\frac{\pi a}{\varepsilon\varepsilon_0\kappa_{{\rm D}}^{2}}\bigg[2\sigma_{{\rm c}}\sigma_{{\rm w}}\ln\left(\frac{\cosh(\kappa_{{\rm D}}d_{0})+\cosh(\kappa_{{\rm D}}x)}{\cosh(\kappa_{{\rm D}}d_{0})-\cosh(\kappa_{{\rm D}}x)}\right)
\\
 & +\left(\sigma_{{\rm c}}^{2}+\sigma_{{\rm w}}^{2}\right)\ln\left(\frac{\cosh(2\kappa_{{\rm D}}d_{0})}{\cosh(2\kappa_{{\rm D}}d_{0})-\cosh(2\kappa_{{\rm D}}x)}\right)\bigg].
 \end{split}
\end{align}
Due to the symmetry, $U'(x)=0$ at $x=0$. However, from Eq.~(\ref{Total_potential}) it follows that this equilibrium is stable (minimum of $U$) only if
\begin{equation} \label{Stability}
\frac{\sigma_{{\rm w}}}{\sigma_{{\rm c}}} <-{\rm e}^{\kappa_{{\rm D}}d_0}~~~~{\rm or}~~~~\frac{\sigma_{{\rm w}}}{\sigma_{{\rm c}}} >-{\rm e}^{-\kappa_{{\rm D}}d_0}.
 \end{equation}
This is exactly the condition that the walls repel the colloid placed at the mid-plane position $x=0$, instead of attracting it~\cite{Parsegian1972}. In what follows, we focus on the case where the colloid and walls have the same charge sign, and the stability condition, Eq.~(\ref{Stability}), is thus always satisfied.

%To leading order, $U$ can be approximated by
Expanding $U(x)$ to second order around the mid-plane, $x=0$, we obtain the harmonic potential approximation
\begin{equation} \label{harmonic}
U\approx \frac{1}{2} K x^2,
\end{equation}
where $K=U''|_{x=0}$ is the effective spring constant. From Eq.~(\ref{Total_potential}), this approximation leads to
\begin{align} \label{spring_constant}
K  = \left( \frac{2\pi a}{\varepsilon_{0}\varepsilon} \right)
\frac{2\sigma_{{\rm c}}\sigma_{{\rm w}}\cosh(\kappa_{{\rm D}}d_{0})+\sigma_{{\rm c}}^{2}+\sigma_{{\rm w}}^{2}}{\sinh^{2}(\kappa_{{\rm D}}d_{0})}.
\end{align}
%%%%%%%%%
%Fig2
\begin{figure}
\includegraphics[width = 0.85 \columnwidth,draft=false]{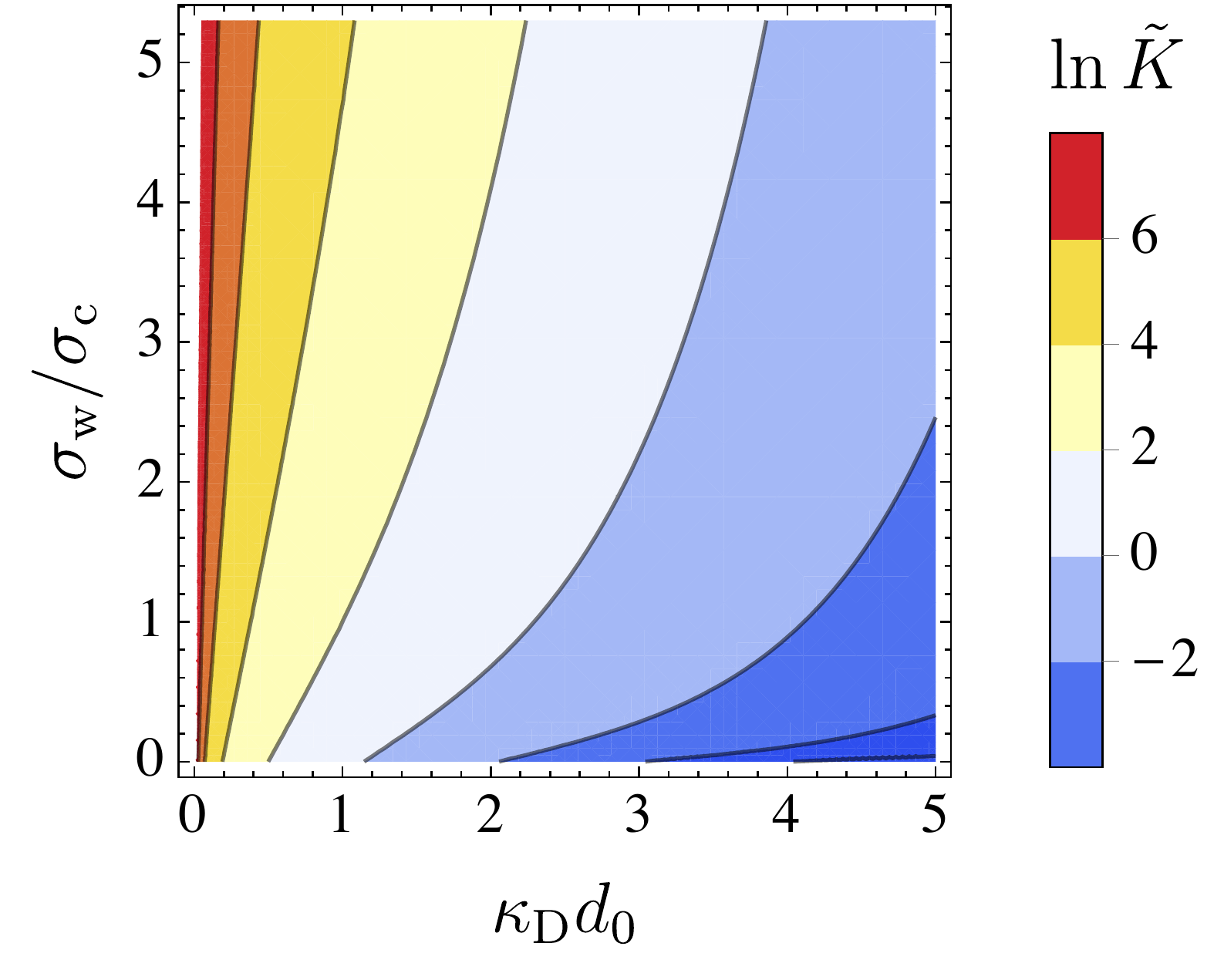} %0.55
\caption{\textsf{Contour plot of the normalized effective spring constant, ${\tilde{K}\equiv K/\left(\pi a \sigma_{\rm c}^2/\varepsilon_0 \varepsilon\right)}$ where $K=U''(x)|_{x=0}$ [see Eqs.~(\ref{harmonic}) and (\ref{spring_constant})] as function of
$\kappa_{\rm D}d_0$ and $\sigma_{\rm w}/\sigma_{\rm c}$. The color code is associated with $\ln\tilde{K}$. Large $\tilde{K}$ is obtained in the high $\sigma_{\rm w}/\sigma_{\rm c}$ and small $\kappa_{\rm D}d_0$ limits. It decays to zero as $\kappa_{\rm D}d_0$ is increased.}}
\label{Fig2}
\end{figure}
%%%%%%%%%%%%%%%%%%%%%%%%%%%%%%%%%%
The harmonic interaction potential in Eq.~(\ref{harmonic}), will be used as the potential energy in the following analysis. The approximation underestimates the strength of the interaction potential close to the walls, but this only has a minor effect on the dynamics, as will be discussed in depth in Sec.~\ref{Discussion}.

In Fig.~\ref{Fig2}, ${\tilde{K}\equiv K/\left(\pi a \sigma_{\rm c}^2/\varepsilon_0 \varepsilon\right)}$ is plotted for different $\kappa_{\rm D} d_0$ and $\sigma_{{\rm w}}/\sigma_{{\rm c}}>0$ values. For fixed $\kappa_{\rm D} d_0$ and $\sigma_{\rm c}$, $K$ diverges as $\sigma_{{\rm w}} \to \infty$, and monotonically decreases as $\sigma_{\rm w}$ decreases. For given $\sigma_{{\rm c}}$ and $\sigma_{{\rm w}}$, $K$ decreases rapidly when $\kappa_{\rm D}d_0$ is increased.

In Appendix~\ref{appendixB} we derive the corresponding $K$ for fixed surface potential on the walls, rather than fixed surface charge.

%%%%
\section{Confined Brownian dynamics} \label{Dynamics}
%%%%
%%%%%%
\subsection{Position-dependent diffusion coefficient (PDDC)}
%%%%%%%
In a free medium (no boundaries), the colloid performs a simple Brownian motion, described by a diffusion coefficient $D$. By the Einstein relation, we have $D=\mu k_{\rm B} T$, where $\mu$ is the mobility, defined by the ratio of the colloid terminal drift velocity to an applied force, $\mu=v/F$.
If the colloid is neutral, and at low Reynolds number, the mobility follows the Stokes' law, $\mu=1/(6\pi \eta a)$, and the diffusion coefficient of this motion is denoted as ${D_{\infty} = k_{\rm B}T/(6\pi\eta a)}$.

The mobility of a charged colloid in an ionic solution is reduced due to the drag of the surrounding ionic cloud. However, for low surface electrostatic potential or low surface charge, the reduction is only of a few percents~\cite{Ohshima1984, Schumacher1987}, and is therefore neglected here. Note that the motion of ions, by virtue of their small size, is much faster than the colloid, and is assumed to be in equilibrium throughout the colloid motion. Their only effect on the colloid dynamics (apart from the reduction of its mobility, which we neglect) is through the equilibrium interaction potential, Eq.~(\ref{harmonic}).

The presence of walls in the vicinity of the colloid, however, can modify the colloid mobility substantially due to hydrodynamic effects. As shown by Brenner~\cite{Brenner1961}, the Stokes' law of a motion in the perpendicular direction to a single solid wall, at distance $d$, is modified in the following way,
\begin{equation} \label{Stokes}
F_\perp=6\pi\eta a\lambda(\zeta) v_\perp,
\end{equation}
with $\lambda$ being
\begin{align} \label{lambda}
\begin{split}
\lambda(\zeta)& =\frac{4}{3}\sinh\zeta\sum_{n=1}^{\infty}\frac{n(n+1)}{(2n-1)(2n+3)}\\
& \times \left[\frac{2\sinh\left[(2n+1)\zeta\right]+(2n+1)\sinh(2\zeta)}{4\sinh^{2}\left[(n+1/2)\zeta \right]-(2n+1)^{2}\sinh^{2}\zeta}-1\right].
\end{split}
\end{align}
Here $\zeta=\cosh^{-1}\left(1+d/a\right)$, and $F_{\perp}$ and $v_{\perp}$ are the force and velocity in the perpendicular direction, respectively. In the $d \gg a$ limit, we have $\lambda =1$ and the regular Stokes' law is recovered. In the $d \ll a$ limit, which is our main interest, Eq.~(\ref{lambda}) becomes, to leading order, $\lambda \approx a/d$.

When the colloid is placed in between two walls, the modified Stokes' law has a much more complicated expression, relying on extensive numerical computations~\cite{Ganatos1980Part1}.
However, several approximations have been proposed~\cite{Faucheux1994,Lobry1996,Benesch2003}. The simplest one is the linear superposition approximation that was shown to agree fairly well with experiments~\cite{Lin2000},
\begin{equation}  \label{Dynamics_eq4}
F_\perp\approx 6\pi \eta a \left(\lambda_{\ell}+\lambda_{{\rm r}}-1\right)v_\perp,
\end{equation}
where $\lambda_{\ell}$ and $\lambda_{{\rm r}}$ are calculated from Eq.~(\ref{lambda}), with $d=d_{\ell}$ and $d=d_{\rm r}$, respectively. Note that the negative 3rd term in Eq.~(\ref{Dynamics_eq4}) guarantees that we recover $F_\perp=6\pi a\eta v_\perp$ when the walls are very far apart.

\begin{figure}
\includegraphics[width = 0.8 \columnwidth,draft=false]{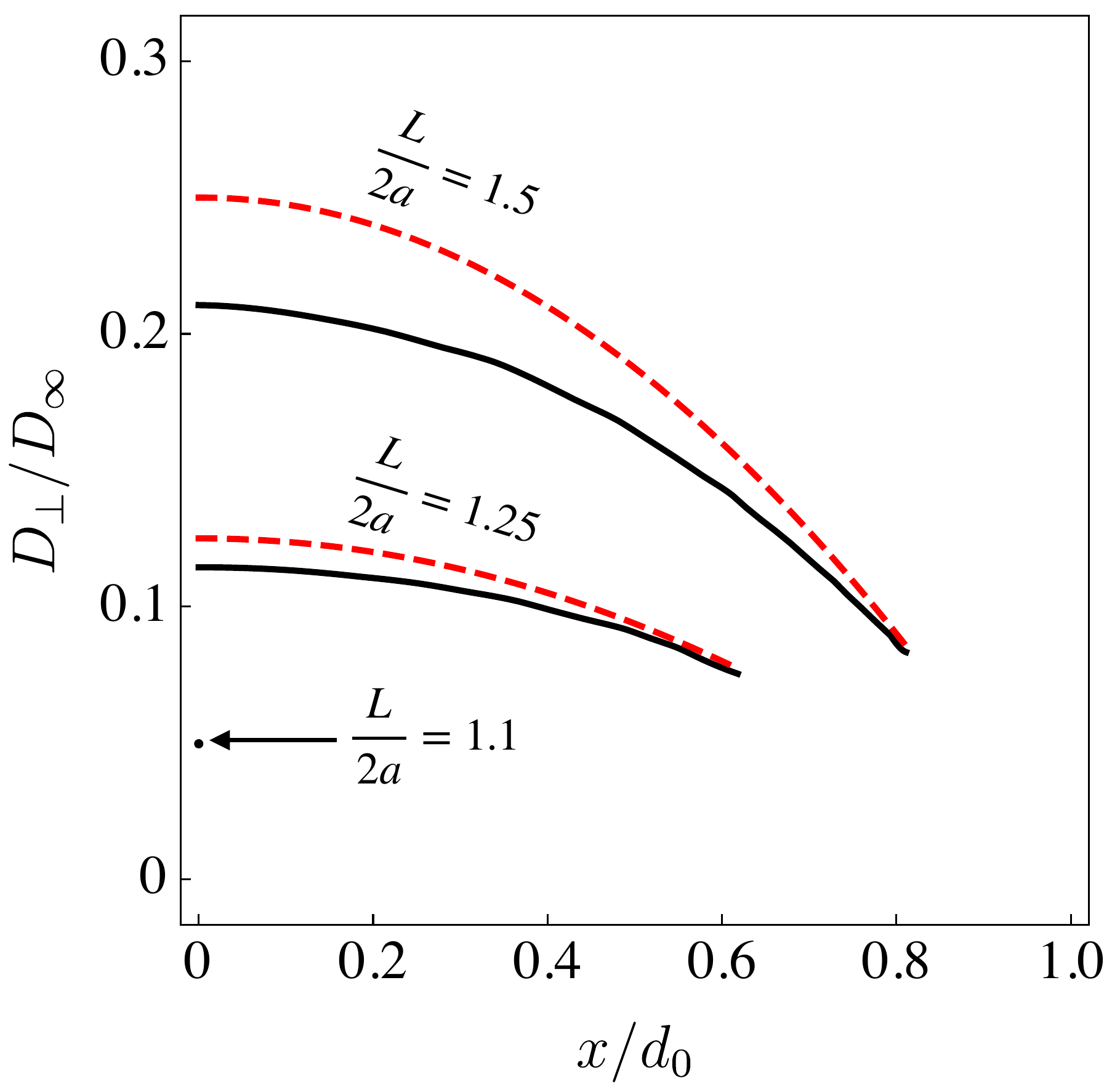}%0.5
\caption{\textsf{The diffusion coefficient $D_\perp$ of a colloid between two flat surfaces, normalized by its diffusion coefficient in a free medium, $D_{\infty} =  k_{\rm B}T/(6\pi\eta a)$, as function of $x/d_0$. Full black lines are numerical results obtained in Ref.~\cite{Ganatos1980Part1}, and red dashed lines show the approximated expression in Eq.~(\ref{Diffusion_Coefficient}), which relies on the linear superposition approximation in the $d_{\ell},d_{\rm r} \ll a$ limit. For $L/(2a)=1.1$, a single data point was available in the numerical calculation: $D_{\perp}/D_{\infty}=0.049$ at $x=0$. The approximated expression in Eq.~(\ref{Diffusion_Coefficient}) yields $D_{\perp}/D_{\infty}(x=0)=0.05$ at $x=0$. The difference between the two values is smaller than the plot resolution.}}
\label{Fig3}
\end{figure}

In the narrow confinement limit, $d_{\ell}$, $d_{\rm r}\ll a$, the perpendicular diffusion coefficient $D_\perp$ can be approximated as
\begin{equation} \label{Diffusion_Coefficient}
D_\perp (x) \approx D_{\infty}\frac{d_{\ell}d_{{\rm r}}}{a\left(d_{\ell}+d_{{\rm r}}\right)}
= D_{0}\left(1-\frac{x^{2}}{d_0^2}\right),
\end{equation}
where
\begin{equation} \label{D0}
D_0=\frac{D_{\infty}d_0}{2a}
\end{equation}
and the relations $d_{\ell}=d_0+x$ and $d_{\rm r}=d_0-x$ were used.
In Fig.~\ref{Fig3}, we compare between the approximated diffusion coefficient given by Eq.~(\ref{Diffusion_Coefficient}), and the ``exact" numerical results of the diffusion coefficient of a colloid between two walls, for motion in the perpendicular direction. The numerical data is adapted from Fig.~6 of Ref.~\cite{Ganatos1980Part1}. Here we show $D_\perp/D_\infty$, which is the inverse of $\lambda$, $D_\perp/D_\infty=1/\lambda$, whereas in Ref.~\cite{Ganatos1980Part1}, $\lambda$ was calculated [see its definition in Eq.~(\ref{Stokes})].

As seen in Fig.~\ref{Fig3}, for $L/(2a)=1.5$, the approximated expression overestimates the ``exact" one by $\sim 20\%$ at the mid-plane ($x=0$), and becomes more accurate as we look away from the mid-plane. For $L/(2a)=1.25$, the deviation in the mid-plane is of $\sim10\%$, and decreasing away from the mid-plane. For $L/(2a)=1.1$, there is only one data point available at the mid-plane, and it deviates from the approximation by only $\sim2\%$. This strongly supports that in the $d_0\ll a$ limit ($L\approx 2a$), the approximated Eq.~(\ref{Diffusion_Coefficient}) can by used.

%%%%%%%%%
%Fig3
%%%%%%%%%%%%%%%%%%%%%%%%%%%%%%%%%%%

The parallel motion is also described by a position dependent diffusion coefficient $D_{\parallel}(x)$ although we do not calculate it here~\cite{Ganatos1980Part2}.

%%%%%%%%%
\subsection{Fokker-Planck equation}
%%%%%%%%%%
The dynamics of a colloid with diffusion coefficients $D_\perp(x)$ and $D_\parallel(x)$ in the perpendicular and parallel directions, respectively, and under potential field $U(x)$, is governed by the generalized Fokker-Planck equation~\cite{Lubensky2007},
\begin{align}\label{Fokker_Planck1}
\begin{split}
\frac{\partial P ({\bf r},t)}{\partial t}&=\frac{\partial}{\partial x}\left[D_\perp(x)\left(\frac{1}{k_{\rm B}T}\frac{\partial U}{\partial x}+\frac{\partial}{\partial x}\right)P({\bf r},t)\right]\\
&+D_\parallel(x)\left(\frac{\partial^2}{\partial y^2}+\frac{\partial^2}{\partial z^2}\right)P\left({\bf r},t\right),
\end{split}
\end{align}
where $P({\bf r},t)$ is the time dependent probability distribution function (PDF) of the colloid position. The full 3D motion of the colloid is quite complex, because the motion in the $y$-$z$ plane is coupled to the motion along the $x$-axis through the diffusion coefficient $D_{\parallel}(x)$. However, our focus is on the perpendicular motion in the $x$-direction, which by symmetry, does not depend on the position in the $y$-$z$ plane. Then one can integrate the equation both over $y$ and $z$ from $-\infty$ to $\infty$, and obtain
\begin{equation}\label{Fokker-Planck_2}
\frac{\partial p (x,t)}{\partial t}=\frac{\partial}{\partial x}\left[D_\perp(x)\left(\frac{1}{k_{\rm B}T}\frac{\partial U}{\partial x}+\frac{\partial}{\partial x}\right)p(x,t)\right],
\end{equation}
where $p(x,t)$ is the reduced probability distribution function $p(x,t)=\int {\rm d}y{\rm d}z \, P({\bf r},t)$.

Substituting the harmonic interaction potential obtained in Eq.~(\ref{harmonic}) and the PDDC of Eq.~(\ref{Diffusion_Coefficient}), we obtain
\begin{equation} \label{Fokker_Planck}
\frac{\partial p}{\partial t}  =D_{0}\frac{\partial}{\partial x}
\left[
\left(1-\frac{x^{2}}{d_{0}^{2}}\right)\left(\frac{Kx}{k_{{\rm B}}T}p+\frac{\partial p}{\partial x}\right)
\right].
\end{equation}
All the electrostatic effects are captured by the term proportional to $K$. We note that, in general, rigid walls impose zero current boundary conditions, i.e., $j=0$ on the walls, where the current is defined by the continuity equation $\partial p/\partial t=-\partial j/\partial x$.
From Eq.~(\ref{Fokker-Planck_2}), the current becomes
$j=-D_\perp(x)\left[(K x/k_{\rm B}T)p+\partial p/\partial x\right]$.
Notice that the conditions $j=0$ at the walls are automatically satisfied by the fact that $D_\perp$ vanishes there.

%%%%%%%%%%%%%%%
\section{Mean square displacement} \label{Results}
%%%%%%%%%%%%%%
Equation~(\ref{Fokker_Planck}), together with the definitions of $K$ and $D_0$ in Eqs.~(\ref{spring_constant}) and~(\ref{D0}), respectively, is the principal equation of this paper. By solving it, one can derive the mean square displacement (MSD), which can be measured in experiments. The MSD of an ensemble of colloids in equilibrium is,
\begin{equation} \label{MSD}
\begin{split}
\langle\left(x(t)-x_{0}\right)^{2}\rangle & =\int_{-d_{0}}^{d_{0}}{\rm d}x_{0}\int_{-d_{0}}^{d_{0}}{\rm d}x\left(x(t)-x_{0}\right)^{2}p\left(x,x_{0};t\right),
\end{split}
\end{equation}
where $p(x,x_0;t)$ is the probability of a colloid to be at $x_0$ at time $t=0$ and at $x$ at time $t$. As the initial position, $x_0$, is drawn from an equilibrium distribution, we can write $p(x,x_0;t)$ in terms of the conditional probability, ${p\left(x,x_0;t\right)=p\left(x,t|x_{0},0\right)p_{{\rm eq}}\left(x_{0}\right)}$, where ${p_{{\rm eq}}\left(x_{0}\right)\propto{\rm exp}({-Kx_{0}^{2}/2k_{{\rm B}}T})}$. It then follows that the MSD is
\begin{equation} \label{MSD}
\begin{split}\langle\left(x(t)-x_{0}\right)^{2}\rangle & =\frac{{\displaystyle \int_{-d_{0}}^{d_{0}}{\rm d}x_{0}\,{\rm e}^{-Kx_{0}^{2}/2k_{{\rm B}}T}\langle\left(x(t)-x_{0}\right)^{2}\rangle_{x_{0}}}}{{\displaystyle \int_{-d_{0}}^{d_{0}}{\rm d}x_{0}\,{\rm e}^{-Kx_{0}^{2}/2k_{{\rm B}}T}}}.
\end{split}
\end{equation}
where $\langle {\cal O} \rangle_{x_{0}}$ is the average over an ensemble of colloids with the same initial position, $x_0$, at $t=0$, {\it i.e}, ${\langle {\cal O} \rangle_{x_{0}}= \int{\rm d}x\, p(x,t|x_{0},0)} {\cal O}$, with ${p(x,t|x_{0},0)}$ being a solution of Eq.~(\ref{Fokker_Planck}) with the initial condition ${p(x,0)=\delta(x-x_0)}$. The MSD resulting from Eq.~(\ref{Fokker_Planck}) will be analyzed in depth in what follows.

%%%%%%%%%%%%%%%%
\subsection{Vanishing  interaction potential limit}
%%%%%%%%%%%%%%
While Eq.~(\ref{Fokker_Planck}) does not have an analytic solution, it can be solved for certain limits. In the {\it vanishing interaction potential limit}, $K$ is omitted altogether and we are left with
\begin{equation} \label{FP1}
\frac{\partial p}{\partial t}=  D_{0}\left[-\frac{2x}{d_{0}^{2}}\frac{\partial p}{\partial x}+\left(1-\frac{x^{2}}{d_{0}^{2}}\right)\frac{\partial^{2}p}{\partial x^{2}}\right].
\end{equation}
The above equation, for a colloid at $x=x_0$ at time $t=0$, has the solution~\cite{Lubensky2007},
\begin{equation} \label{NoPotential}
\begin{split}
p(x,t|x_0,0) & =\frac{1}{2d_{0}}\sum_{n=0}^{\infty}(2n+1){\rm e}^{-D_{0}n(n+1)t/d_0^2} \\
& \times P_{n}\left(\frac{x_{0}}{d_{0}}\right)P_{n}\left(\frac{x}{d_{0}}\right),
\end{split}
\end{equation}
where $P_n(x)$ is the Legendre polynomial of order $n$. The PDF of Eq.~(\ref{NoPotential}), is characterized by the following first and second moments,
\begin{equation} \label{Dyn_System2_eq10}
\begin{split}
\langle x(t)\rangle_{x_{0}} & =x_{0} {\rm e}^{-2D_{0}t/d_{0}^{2}},\\
\langle x^{2}(t)\rangle_{x_{0}} & =x_{0}^{2} {\rm e}^{-6D_{0}t/d_{0}^{2}}+\frac{d_{0}^{2}}{3}\left(1-{\rm e}^{-6D_{0}t/d_{0}^{2}}\right),
\end{split}
\end{equation}
In the limit of $t\to \infty$, the PDF reduces to a uniform distribution between the two walls
that are positioned at ${x=\pm d_0}$.

In the vanishing interaction potential limit, $K=0$, Eq.~(\ref{MSD}) simply becomes
\begin{align}
\begin{split}
& \langle\left(x(t)-x_{0}\right)^{2}\rangle \\
& = \frac{1}{2d_{0}}\int_{-d_{0}}^{d_{0}}{\rm d}x_{0} \, \left[\langle x^2(t)\rangle_{x_{0}} \right.
\left. -2x_{0}\langle x(t)\rangle_{x_{0}} +x_{0}^{2}\right],
\end{split}
\end{align}
leading to the MSD
\begin{equation} \label{MSD1}
\langle\left(x(t)-x_{0}\right)^{2}\rangle=\frac{2d_{0}^{2}}{3}\left(1-{\rm e}^{-2D_{0}t/d_{0}^{2}}\right).
\end{equation}
The above MSD approaches $2d_0^2/3$ in the limit of $t\to\infty$.

%%%%%%%%%%%%%%%
\subsection{Strong interaction potential limit}
%%%%%%%%%%%%%%
In the opposite {\it strong interaction potential limit}, the confinement due to electrostatic effects effect is strong, and the colloid remains close to the mid-plane. Consequently, its diffusion coefficient does not change, and equals to $D_{0}$.
Then the term that depends on $d_0$ in Eq.~(\ref{Fokker_Planck}) can be omitted, and we obtain
\begin{equation}\label{FP2}
\begin{split}\frac{\partial p}{\partial t} =D_{0}\left[\frac{K}{k_{{\rm B}}T}\left(p+x\frac{\partial p}{\partial x}\right)+\frac{\partial^{2}p}{\partial x^{2}}\right].
\end{split}
\end{equation}
Note that this is equivalent to the case of permeable walls, where the walls interact electrostatically but not hydrodynamically. Additionally, note that a strong interaction potential $U$ does not contradict the assumption employed in Sec.~\ref{Equilibrium} of a small electrostatic potential $\psi$. This is evident, for example, by the dependence of $U$ on the colloid size (see Eq.~(\ref{Total_potential})), which is absent in $\psi$ (see Eq.~(\ref{elec_potential})).

The above equation can be solved for a colloid at $x=x_0$ at time $t=0$, yielding~\cite{Risken},
\begin{equation} \label{StrongPotential}
\begin{split}
p(x,t|x_0,0)& =\left( \frac{K}{2\pi k_{\rm B}T(1-{\rm e}^{-2 D_{0}Kt/k_{\rm B}T})} \right)^{1/2}
\\
&\times\exp\left[-\frac{K(x-x_{0}{\rm e}^{-D_{0}Kt/k_{\rm B}T})^{2}}
{2k_{\rm B}T(1-{\rm e}^{-2D_{0}Kt/k_{\rm B}T})}\right].
\end{split}
\end{equation}
For the averages, we can extend the integral range in ${\langle {\cal O} \rangle_{x_{0}}= \int{\rm d}x \, p(x,t|x_{0},0)} {\cal O}$ from $\pm d_0$ to $\pm \infty$. Then the first and second moments are
\begin{align} \label{Dyn_System1_eq7}
\begin{split}
\langle x(t)\rangle_{x_0} & = x_{0}{\rm e}^{-D_{0}Kt/k_{{\rm B}}T}, \\
\langle x^2(t)\rangle_{x_0} & = x_{0}^{2}{\rm e}^{-2D_{0}Kt/k_{{\rm B}}T}
\\
& + \frac{k_{{\rm B}}T}{K}\left(1-{\rm e}^{-2D_{0}Kt/k_{{\rm B}}T}\right).
\end{split}
\end{align}
For the MSD, Eq.~(\ref{MSD}) now reads
\begin{equation}
\begin{split}
\langle\left(x(t)-x_{0}\right)^{2}\rangle & =\left( \frac{K}{2\pi k_{{\rm B}}T} \right)^{1/2}
\int_{-\infty}^{\infty} {\rm d}x_{0} \, {\rm e}^{-Kx_{0}^{2}/2k_{{\rm B}}T}\\
&\times\left[\langle x^2(t)\rangle_{x_{0}}-2x_{0}\langle x(t)\rangle_{x_{0}}+x_{0}^{2}\right],
\end{split}
\end{equation}
yielding,
\begin{equation} \label{MSD2}
\langle\left(x(t)-x_{0}\right)^{2}\rangle = \frac{2k_{{\rm B}}T}{K}\left(1-{\rm e}^{-D_{0} K t/k_{{\rm B}}T}\right),
\end{equation}
which asymptotically approaches $2k_{{\rm B}}T/K$ for $t\to\infty$.
Equations~(\ref{NoPotential}) and~(\ref{StrongPotential}) and their corresponding MSD functions, Eqs.~(\ref{MSD1}) and~(\ref{MSD2}), represent the two limiting dynamical behaviors.

%%%%%%%%%%%%
\subsection{Short-time diffusion coefficient}
%%%%%%%%%%%%%
For a known initial position, $p(x,0)$ is infinitely sharp (Dirac delta function). As a result, the term proportional to the second derivative of $p$ in Eq.~(\ref{Fokker_Planck}) dominates the dynamics for short times,
\begin{equation} \label{Short_time}
\frac{\partial p}{\partial t} = D_{\rm eff}(x_0)\frac{\partial^{2}p}{\partial x^{2}}.
\end{equation}
with $D_{\rm eff}(x_0)\equiv D_{0}\left(1-x_{0}^{2}/d_{0}^{2}\right)$. The colloid experiences, for a short while, a free diffusion with diffusion coefficient $D_{\rm eff}(x)$, with the following conditional probability distribution function,
\begin{align}
\begin{split}
p(x,t|x_0,0) =\left( \frac{1}{4\pi D_{\rm eff}(x_0)t} \right)^{1/2} \exp\left[-\frac{(x-x_{0})^{2}}{4D_{\rm eff}(x_0)t}\right].
\end{split}
\end{align}

This results in a linear time dependence of MSD, $\langle\left(x-x_{0}\right)^{2}\rangle=2D_{\rm app}t$,
where $D_{\rm app}$ is the apparent diffusion coefficient
\begin{equation} \label{Diffusion_alpha}
\begin{split}D_{\rm app} & =\frac{\displaystyle \int_{-d_{0}}^{d_{0}}{\rm d}x_{0}\,{\rm e}^{-Kx_{0}^{2}/2k_{{\rm B}}T}D_{\rm eff}(x_0)}{\displaystyle \int_{-d_{0}}^{d_{0}}{\rm d}x_{0}\,{\rm e}^{-Kx_{0}^{2}/2k_{{\rm B}}T}}\\
 & =D_{0}\left[1-\frac{1}{\alpha}+\frac{\sqrt{2/(\pi\alpha)}\,{\rm e}^{-\alpha/2}}{\text{erf}(\sqrt{\alpha/2})}\right],
\end{split}
\end{equation}
where $\alpha$ is a dimensionless parameter defined by
\begin{equation} \label{alpha_def}
\alpha=\frac{Kd_{0}^{2}}{k_{{\rm B}}T}
\end{equation}
and $\text{erf}(x)=\left(2/\sqrt{\pi}\right)\int_{0}^{x}{\rm e}^{-z^{2}} {\rm d}z$ is the error function. Notice that $\sqrt{\alpha}$ is the ratio between two characteristic length scales, $d_0$ and $\sqrt{k_{\rm B}T/K}$. While $d_0$ is the length over which the diffusion coefficient changes, $\sqrt{k_{\rm B}T/K}$
is the characteristic interaction potential length scale, which is small for strong potentials and large for weak ones.

\begin{figure}
\includegraphics[width = 0.9 \columnwidth,draft=false]{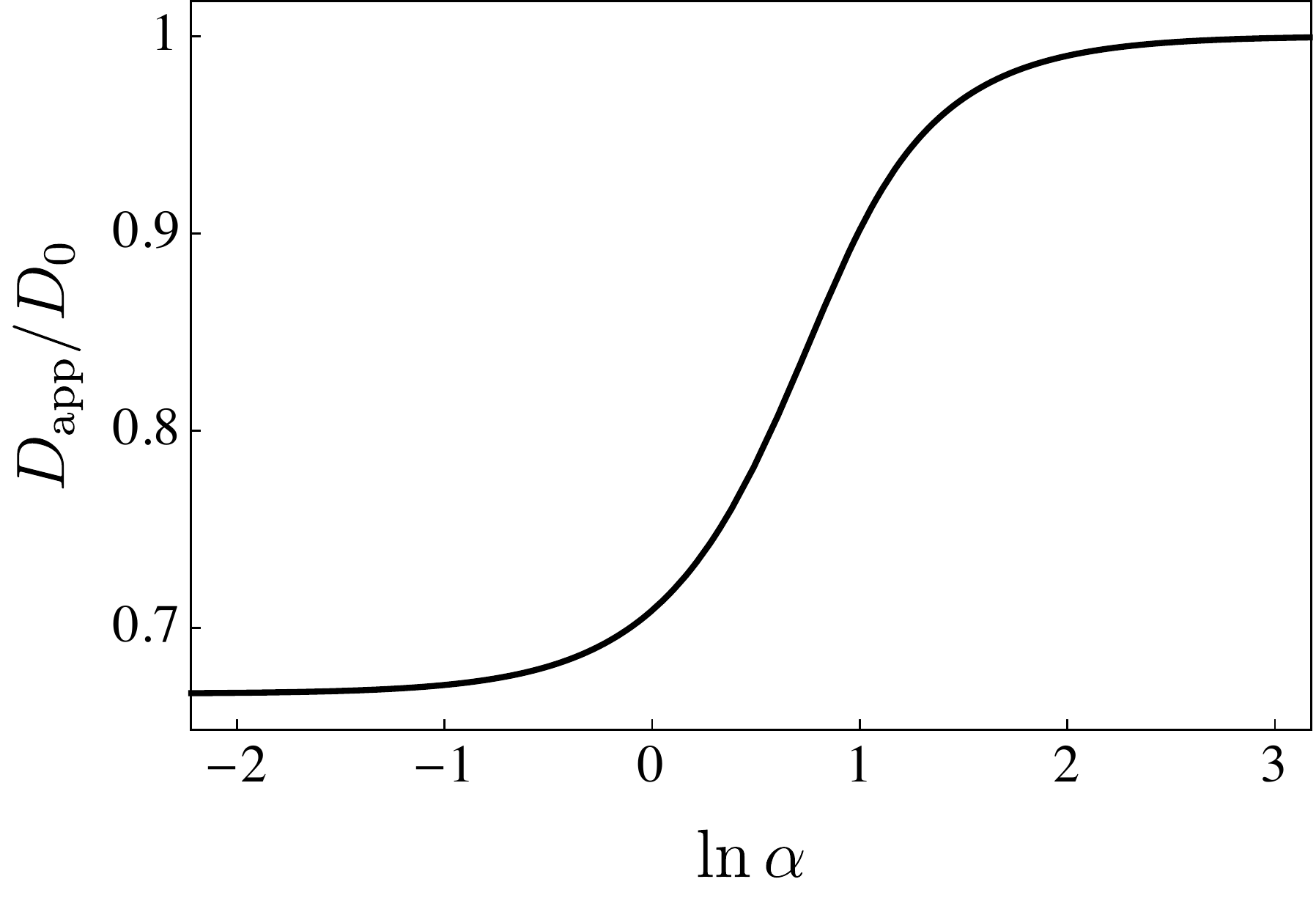}%0.5
\caption{\textsf{The apparent diffusion coefficient $D_{\rm app}$ [see Eq.~(\ref{Diffusion_alpha})], scaled by $D_0$, as a function of $\alpha=Kd_{0}^{2}/k_{{\rm B}}T$.}}
\label{Fig4}
\end{figure}

In Fig.~\ref{Fig4}, $D_{\rm app}/D_0$ is plotted as a function of $\alpha$. Upon increasing $\alpha$, $D_{\rm app}$ monotonically grows from $D_{\rm app}/D_0=2/3$ obtained for $\alpha=0$ to $D_{\rm app}/D_0=1$ valid for the $\alpha\to\infty$ limit. We conclude that the colloid moves faster as the electrostatic effects are stronger (large $\alpha$). The change in $D_{\rm app}$ occurs when $\alpha$ is of order unity.
%%%%%%%%%%%%%%%%%%%%%%%%%%%%%%%%%%%%%%%%%%%
%Fig4
%%%%%%%%%%%%%%%%%%%%%%%%%%%%%%%%%%%%%%%

%%%%%%%%%%%%%%%%%%%%%%%%%%%%%%%%%%%%%%%%%%%
%Fig5
\begin{figure*}
\includegraphics[width = 1.6 \columnwidth,draft=false]{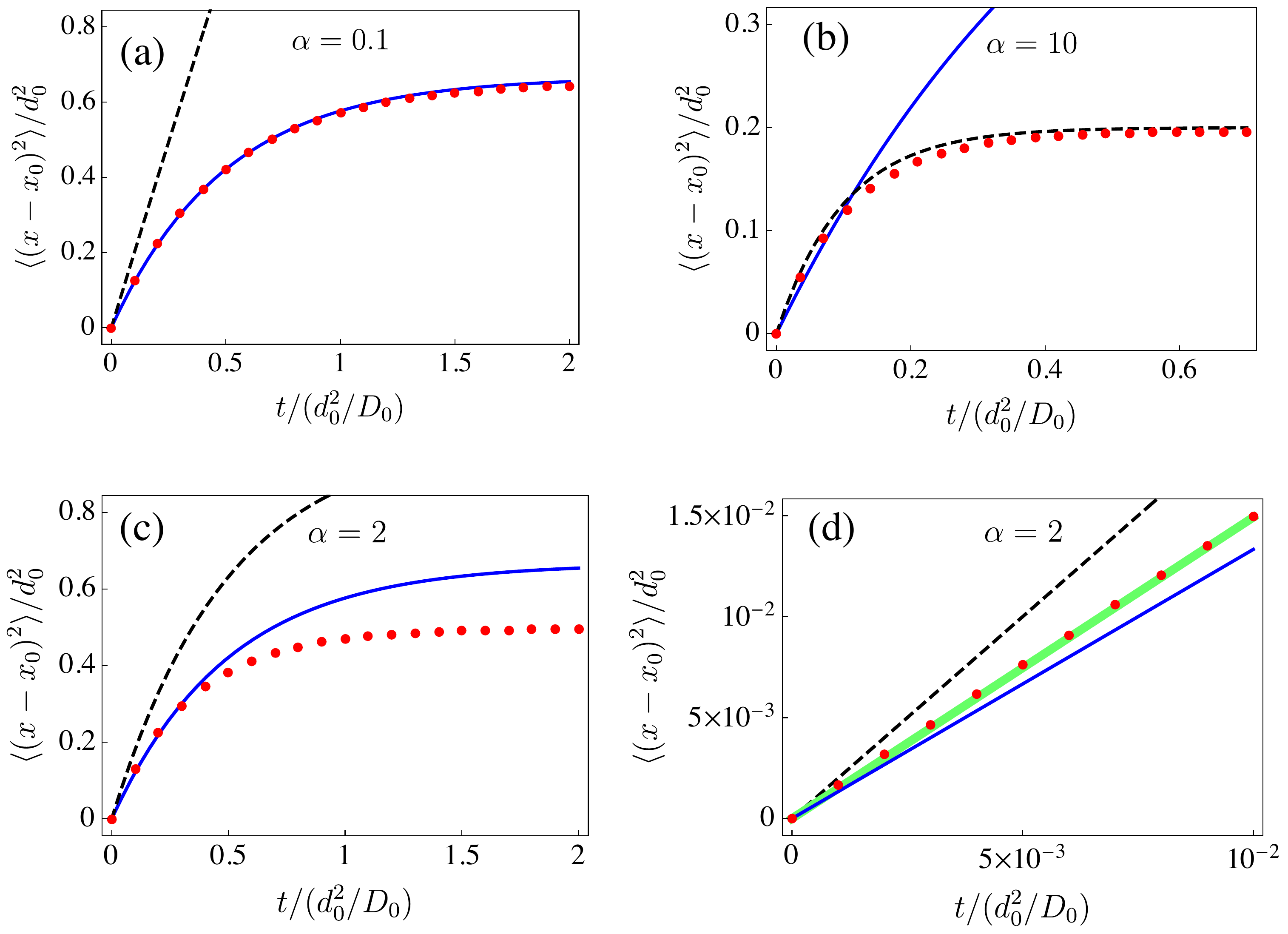}%1.1
\caption{\textsf{The mean square displacement (MSD) of the colloid, $\langle\left(x-x_0\right)^2 \rangle$ scaled by $d_0^2$, as a function of $t$ scaled by $d_0^2/D_0$ for (a) $\alpha=0.1$, (b) $\alpha=10$, (c) and (d) $\alpha=2$. Red dots are the numerical solution of Eq.~(\ref{Fokker_Planck}). Solid blue line is the vanishing interaction potential limit [see Eq.~(\ref{MSD1})]. Dashed black line is the strong interaction potential limit [see Eq.~(\ref{MSD2})]. In (d), the short-time limit of $\alpha=2$ is shown, together with the prediction, $\langle\left(x-x_{0}\right)^{2}\rangle=2D_{\rm app}t$ with $D_{\rm app}$ defined in Eq.~(\ref{Diffusion_alpha}) (thick green line).}}
\label{Fig5}
\end{figure*}
%%%%%%%%%%%%%%%%%%%%%%%%%%%%%%%%%%%%%%%

%%%%%%%%%%%%
\subsection{Numerical analysis} \label{general}
%%%%%%%%%%%%%

The full dynamical equation, Eq.~(\ref{Fokker_Planck}), can be made dimensionless by using the dimensionless
position $\tilde{x}=x/d_0$ and time $\tilde{t}=(D_{0}/d_0^2)t$ as
\begin{equation} \label{FP_normalized}
\frac{\partial p}{\partial\tilde{t}} =\frac{\partial}{\partial\tilde{x}}(1-\tilde{x}^{2})\left(\alpha\tilde{x}p+\frac{\partial p}{\partial\tilde{x}}\right),
\end{equation}
where $\alpha$ is given by Eq.~(\ref{alpha_def}). The vanishing interaction potential limit, Eq.~(\ref{FP1}), and strong interaction potential limit, Eq.~(\ref{FP2}), correspond to $\alpha = 0$ and to the limit $\alpha \to \infty$, respectively (in order to obtain Eq.~(\ref{FP2})  from Eq.~(\ref{FP_normalized}), one needs to redefine the normalized position and time).

We calculate the MSD of the particle for different $\alpha$ values by solving Eq.~(\ref{FP_normalized}) numerically. In the following we use the same normalized variables as in Eq.~(\ref{FP_normalized}), but drop the tilde signs for brevity. We divide our space into $N$ lattice points $\{x_1,...,x_N\}$, where $x_1=-1$ and $x_N=1$, and define $p_i(t)\equiv p(x_i,t)$. For a given $x_0$, we start with a distribution that approximates a Dirac delta function, $p_i(0)=(2\pi \delta)^{-1/2}\exp[(x_i-x_0)^2/2\delta]$, with $\delta \ll 1$. We then iterate each time step using the Euler method,
\begin{align}
\begin{split}
& p_{i}(t+\Delta t) =p_{i}(t)\\
& +\frac{\Delta t}{\Delta x} \bigg[(1-x_{i+1}^{2})\left(\alpha x_{i+1} p_{i+1}(t)+\frac{p_{i+1}(t)-p_{i}(t)}{\Delta x}\right)\\
& -(1-x_{i}^{2})\left(\alpha x_{i} p_{i}(t)+\frac{p_{i}(t)-p_{i-1}(t)}{\Delta x}\right)\bigg].
\end{split}
\end{align}
The zero current boundary condition is guaranteed due to the cancelation of the diffusion coefficient, proportional to $(1-x_i^2)$, at $x_1$ and $x_N$. We used a lattice spacing of $\Delta x=0.02$, and a time step $\Delta t$ between $6.25\times 10^{-5}$ and $2\times 10^{-4}$, depending on $\alpha$.
For each initial condition, $x_0 \in \{-1,1\}$, we calculated $\langle\left(x(t)-x_{0}\right)^{2}\rangle_{x_0}$ and obtained the MSD through Eq.~(\ref{MSD}).

The numerically obtained full MSD is shown in Fig.~\ref{Fig5} for different $\alpha$ values, and is compared with the limits in Eqs.~(\ref{MSD1}) and~(\ref{MSD2}). Figure~\ref{Fig5}(a) shows that for $\alpha=0.1$, the MSD coincides, almost perfectly, with the vanishing interaction potential case (solid blue line), while in Fig.~\ref{Fig5}(b) for $\alpha=10$, it coincides with the strong interaction potential case (dashed black line). However, for an intermediate value, $\alpha=2$ in Fig.~\ref{Fig5}(c), the numerically obtained MSD deviates substantially from the two limits. In the short-time limit, it lies between the two limiting cases, while in the long-time limit it has a significantly lower value than both limits.
In the short-time region shown in Fig.~\ref{Fig5}(d), the numerical calculation coincides with Eq.~(\ref{Diffusion_alpha})
(thick green line), and the MSD grows linearly with a slope that increases with $\alpha$. At saturation, the MSD  for $\alpha=0.1$ equals $\sim0.6 d_0^2$, for $\alpha=2$ equals $\sim0.4 d_0^2$ and for $\alpha=10$, it approaches $\sim0.2 d_0^2$, indicating that the colloid motion is more localized when the electrostatic force is strong.

\begin{figure}
\includegraphics[width = 0.9 \columnwidth,draft=false]{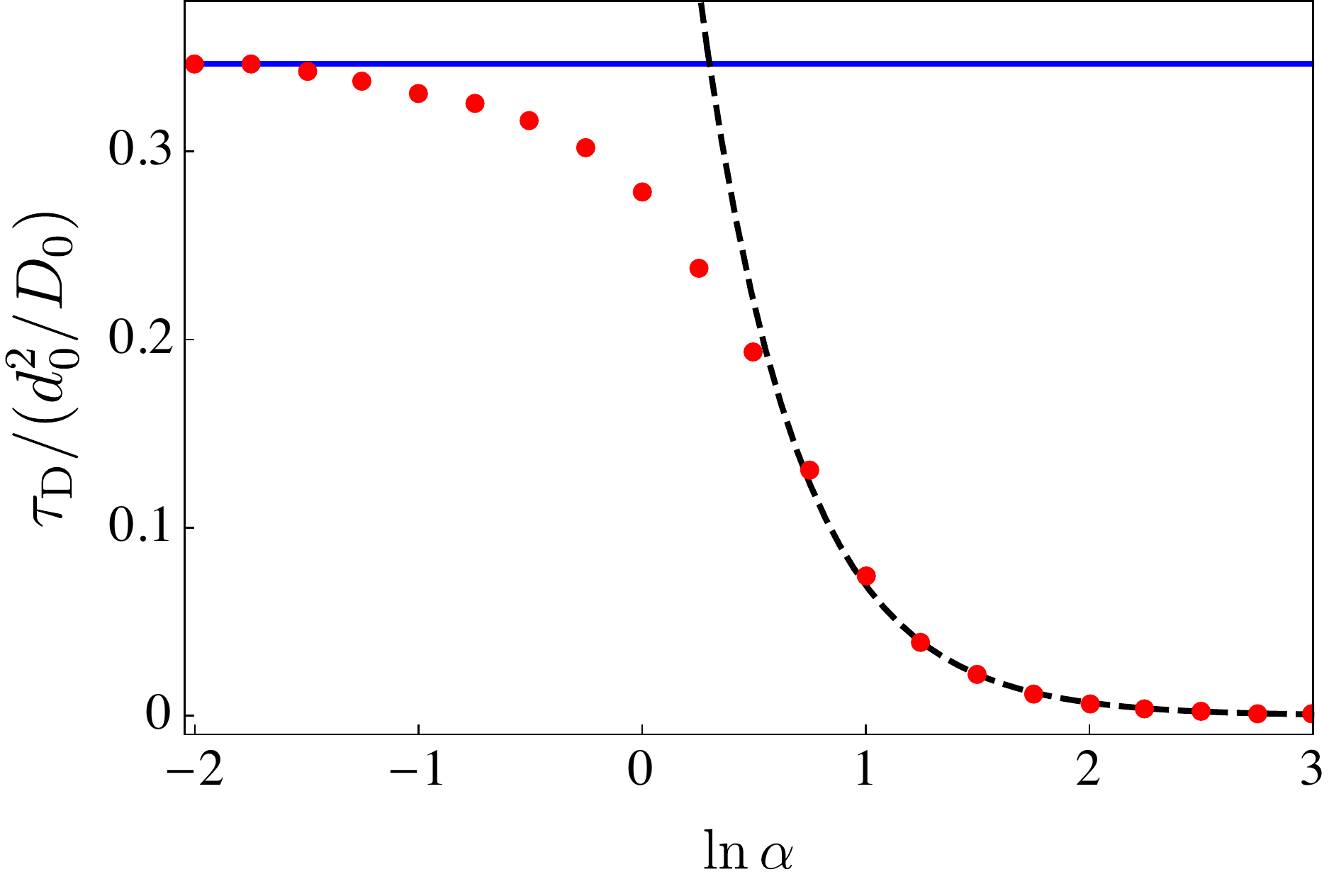}%0.5
\caption{\textsf{The diffusion characteristic time $\tau_{\rm D}$, defined as the time at which the MSD reaches half of its value in the $t\to\infty$ limit, normalized by $d_0^2/D_0$, as a function of $\alpha$. Solid blue line is the vanishing interaction potential limit, $\tau_{\rm D}=d_0^2\ln 2/(2D_0)$, and dashed black line is the strong interaction potential walls limit, $\tau_{\rm D}=d_0^2\ln2/(D_0 \alpha)$. Red dots are the values calculated numerically from Eq.~(\ref{Fokker_Planck}) and interpolate between the two limits.}}
\label{Fig6}
\end{figure}

A distinct difference between the MSD for different $\alpha$ values is the characteristic time at which MSD saturates.
This is the time it takes the colloid to explore its available space.
Denoting $\tau_{\rm D}$ as the time at which the MSD reaches half of its maximal value (the diffusion characteristic time), we obtain that in the $\alpha=0$ case, $\tau_{\rm D}=d_0^2\ln2/(2D_0)$ [see Eq.~(\ref{MSD1})], while in the $\alpha \to \infty$ limit,
$\tau_{\rm D}=k_{\rm B} T \ln2/(D_0 K)=d_0^2\ln2/(D_0 \alpha)$ [see Eq.~(\ref{MSD2})].
In Fig.~\ref{Fig6}, $\tau_{\rm D}$ scaled by  $d_0^2/D_0$,  is shown as a function of $\alpha$.
Similar to $D_{\rm app}$, the diffusion characteristic time $\tau_{\rm D}$ interpolates between the two limits mentioned above, where the
crossover occurs around $\alpha \sim 1$. The decrease of $\tau_{\rm D}$ as function of $\alpha$, seen in Fig.~\ref{Fig6}, is attributed to two effects. First, as $\alpha$ increases, the colloid is more localized due to electrostatic forces and explores a smaller region, and second, its average diffusion coefficient, being further away from the walls, becomes larger.

%%%%%%%%
%Fig6
%%%%%%%%%%%%%%%%%%%%%%%%%%%%%%%%%%%%%%%%%%%

%%%%%%%%%
\subsection{Dependence of $\alpha$ on electrostatic properties}
\begin{figure}
\includegraphics[width = 0.9 \columnwidth,draft=false]{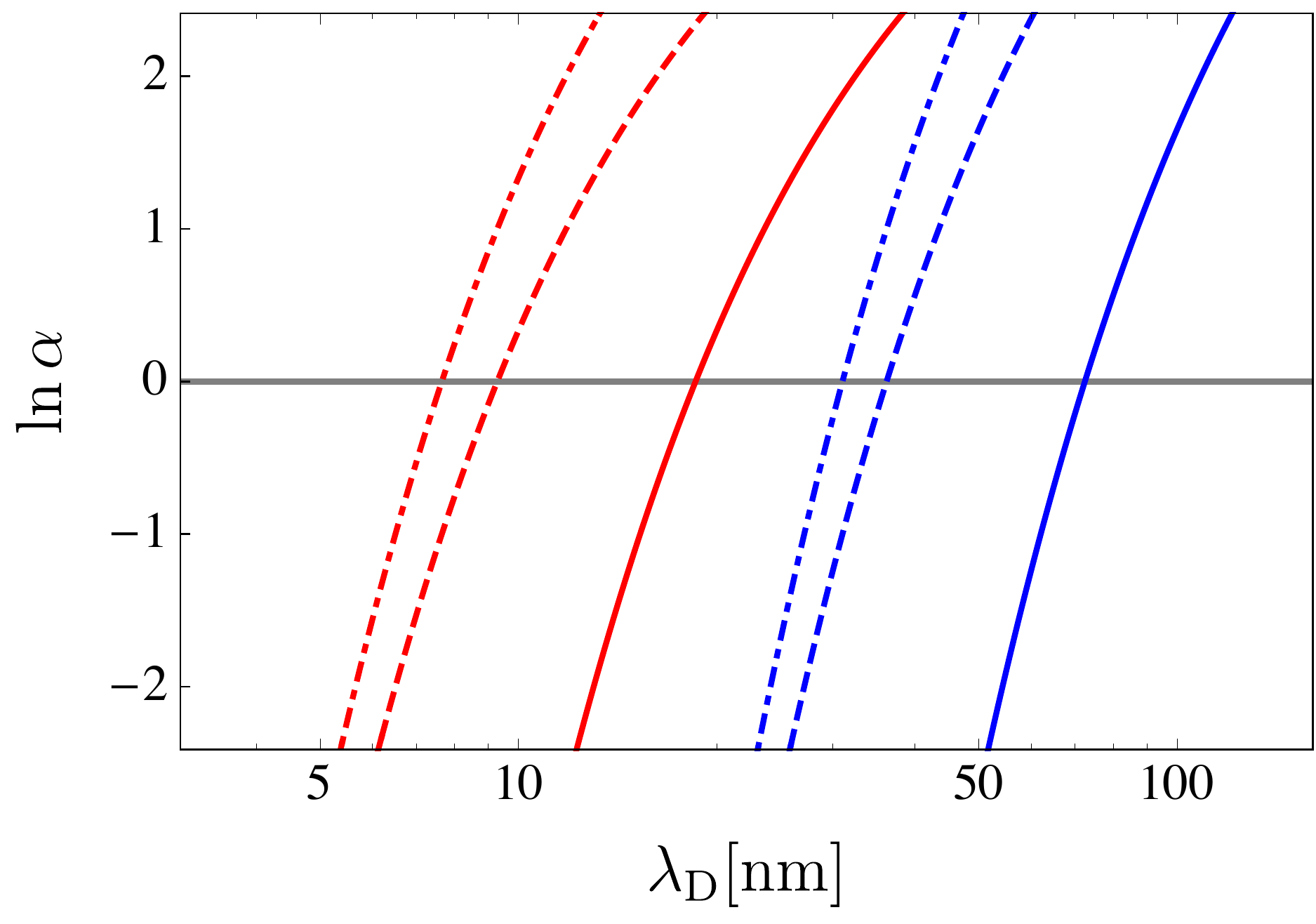}%0.5
\caption{\textsf{The natural logarithm of the dimensionless parameter $\alpha=Kd_{0}^{2}/k_{{\rm B}}T$, where $K$ is given by Eq.~(\ref{spring_constant}), as a function of the screening length, $\lambda_{\rm D}$ on a logarithmic scale.
The system parameters are:
$L=5\,{\rm \mu m}$,
$\sigma_{\rm c}=0.003\, e/{\rm nm}^{2}$,
$T=300\,{\rm K}$ and
$\varepsilon=80$.
Blue lines are for $d_0=500\,{\rm nm}$ ($a=2.0\,{\rm \mu m}$), and red lines are for $d_0=100\,{\rm nm}$ ($a=2.4\,{\rm \mu m}$). Solid lines are for uncharged walls, $\sigma_{\rm w}=0$, dashed lines are for walls with surface charge density $\sigma_{\rm w}=\sigma_{\rm c}$, and dotted-dashed line are for $\sigma_{\rm w}=10\sigma_{\rm c}$.}}
\label{Fig7}
\end{figure}

%%%%%%%%%
So far we have seen that the dynamics are determined by the value of $\alpha$. Substituting the value of $K$ derived in Eq.~(\ref{spring_constant}), we obtain
\begin{equation} \label{alpha}
\alpha=\left( \frac{2\pi ad_{0}^{2}}{\varepsilon_{0}\varepsilon k_{\rm B}T} \right)
\frac{2\sigma_{{\rm c}}\sigma_{{\rm w}}\cosh(\kappa_{{\rm D}}d_{0})+\sigma_{{\rm c}}^{2}+\sigma_{{\rm w}}^{2}}{\sinh^{2}(\kappa_{{\rm D}}d_{0})}.
\end{equation}
In an experimental setup, $\alpha$ can be controlled by changing the ionic strength (salt concentration, $n$), and consequently the Debye length, $\lambda_{\rm D}\sim1/\sqrt{n}$. The dependence of $\alpha$ on $\lambda_{\rm D}=\kappa^{-1}_{\rm D}$ is plotted in Fig.~\ref{Fig7}, for reasonable colloid and solvent parameters. As expected, $\alpha$ is large for large screening length and decreases towards zero as $\lambda_{\rm D}$ decreases.

As we have shown, $\alpha=1$ signifies the crossover from a PCCD dominated motion to an electrostatically dominated one. For the range of parameters in Fig.~\ref{Fig7}, this crossover occurs roughly when $\lambda_{\rm D}\sim d_0/10$. The screening length that corresponds to the crossover increases with $d_0$, and for a fixed $\sigma_{\rm c}$ ($\sigma_{\rm w}$), it decreases as $\sigma_{\rm w}$ ($\sigma_{\rm c}$) increases.

%%%%%%%%%
%Fig7
%%%%%%%%%%%%%%%%%%%%%%%%%%%%%%%%%%%

%%%%%%%%%%%%
\section{Summary and discussion} \label{Discussion}
%%%%%%%%%%%%
We studied the dynamics of a charged colloid under restricted confinement of two charged surfaces. The combination of electrostatic and hydrodynamic forces exerted by the walls result in a unique behavior. This behavior can be quantified in term of a dimensionless parameter $\alpha$ in Eq.~(\ref{alpha_def}) that determines the interplay between the electrostatic interaction, and the position dependent drag force. The parameter $\alpha$ is also given by Eq.~(\ref{alpha}), showing how the colloid motion, including its short-time behavior, Eq.~(\ref{Diffusion_alpha}) and long-time behavior (Figs.~\ref{Fig5} and~\ref{Fig6}), depends on the geometry and electrostatic properties. In particular, $\alpha$ can be tuned by changing the screening length (Fig.~\ref{Fig7}) that is usually an easily controlled parameter in experiments. At small screening length (small $\alpha$), the colloid moves slower than at larger screening length (large $\alpha$). In addition, as $\lambda_{\rm D}$ is increased the colloid explores a smaller region in space throughout its motion. The two effects lead to a decrease in the diffusion characteristic time, $\tau_{\rm D}$, as $\lambda_{\rm D}$ is increased.

We note that the harmonic approximation of the interaction potential allows us to obtain analytical results. If we use the full interaction potential, Eq.~(\ref{Total_potential}), in the dynamical equation, we could not have expressed it in terms of a single parameter, $\alpha$, as in Eq.~(\ref{FP_normalized}). This harmonic approximation underestimates the strength of the interaction potential near the walls. However, as long as $d_0$ is of the order of the screening length, or smaller, the approximation is valid except for the regime very close to the wall where the interaction potential $U$ diverges. This small region near the walls does not affect the MSD that integrates the motion throughout the entire space.
Moreover, the divergence of $U$ close to the wall is unphysical because at such close proximity the DH theory, Eq.~(\ref{DH}), is no longer valid. For $d_0\gg \lambda_{\rm D}$,
the following limiting expression should be used instead
\begin{equation} \label{lim_U}
U\left(x\right)\approx\frac{2Q\sigma_{{\rm w}}\,}{\varepsilon_{0}\varepsilon\kappa_{{\rm D}}^{2}a}{\rm e}^{-\kappa_{{\rm D}}d_0}\cosh(\kappa_{{\rm D}}x).
\end{equation}
However, as in this limit the interaction potential is very small and can be ignored, there is no substantial difference in the MSD between the exact and approximated potentials.

We also note that, in our analysis, the effect of the ionic cloud on the diffusion coefficient of the charged colloid was neglected. In the free medium, the effect of the ionic cloud was calculated and turned out to be very small (of only a few percents)~\cite{Ohshima1984, Schumacher1987}. Since in a restricting confinement this effect was not fully investigated~\cite{Eichmann2008,Chun2004}, it is not considered here for simplicity.

Although the above mentioned calculations are approximated, we expect our results to qualitatively describe the
Brownian motion of a charged colloid between confining charged walls.
This should be tested in future experiments that will focus on very restricted confinement, as considered here.

%%%%%%%%%%%
\acknowledgements
%%%%%%%%%%%

We would like to thank R. Adar for useful discussions. Y.A.\ is thankful for the support of the Clore Scholars Programme of the Clore Israel Foundation and
the hospitality at Tokyo Metropolitan University, where part of this work has been conducted.
S.K.\ acknowledges support by a Grant-in-Aid for Scientific Research (C) (Grant No.\ 18K03567 and Grant No.\ 19K03765) from the Japan Society for the Promotion of Science and support by a Grant-in-Aid for Scientific Research on Innovative Areas ``Information Physics of Living Matters'' (Grant No.\ 20H05538) from the Ministry of Education, Culture, Sports, Science and Technology of Japan. D.A.\ acknowledges support from the Israel Science Foundation (ISF) under Grant No.\ 213/19.

\appendix
%%%%%%%%%%%
%%%%%%%%%
\section{Derivation of Eq.~(\ref{Total_potential})} \label{appendixA}
%%%%%%%%%%%%%
In the large sphere limit, the force acting on the colloid from the right wall is independent of that acting from the left wall, and vise versa. For each side, the force on the colloid is approximated by the Derjaguin approximation~\cite{Israelachvili1992}
\begin{equation} \label{Force}
F\left(d\right)\approx  2\pi a\int_{d}^{\infty}{\rm d}l\,\Pi(l),
\end{equation}
where $\Pi$ is the force per unit area exerted on the colloid, if the colloid is considered as flat rather than spherical, and $d$ is the distance between the colloid and the wall ($d=d_{\ell}$ for the left wall and $d=d_{\rm r}$ for the right wall).

The electrostatic potential in the Derjaguin approximation can be calculated in a straightforward way. As it was performed multiple times for other geometries and boundary conditions~\cite{Israelachvili1992}, we describe it only briefly here. Since the two sides of the colloid are decoupled, we start by solving the DH equation in Eq.~(\ref{DH}) for a single wall, while assuming that the colloid is flat ($a\to\infty$). Denoting the distance from the left wall, without loss of generality, by $\xi$, the boundary conditions are ${\rm{d}\psi/\rm{d}\xi|_{\xi=0}=-\sigma_{\rm w}/\left(\varepsilon_0\varepsilon\right)}$ and ${\rm{d}\psi/\rm{d}\xi|_{\xi=d} = \sigma_{\rm c}/\left(\varepsilon_0\varepsilon\right)}$. Then the solution is
\begin{equation} \label{elec_potential}
\psi \left(\xi;d\right)=\frac{\sigma_{{\rm w}}\cosh\left[\kappa_{{\rm D}}\left(d- \xi\right)\right]+\sigma_{{\rm c}}\cosh (\kappa_{{\rm D}}\xi )}{\varepsilon\varepsilon_0\kappa_{{\rm D}}\sinh\left(\kappa_{{\rm D}}d\right)}.
\end{equation}

The DH approximation is valid when the electrostatic potential on the wall ($\xi=0$) and on the colloid surface ($\xi=d$) is small compared to $k_{\rm B}T/e$~\cite{Tomer2020}. This is satisfied when $\sigma_{{\rm c}},\sigma_{{\rm w}}\ll e/(\ell_{{\rm B}}\lambda_{{\rm D}})$ and $\sigma_{{\rm c}},\sigma_{{\rm w}}\ll e d/\left(l_{{\rm B}}\lambda_{{\rm D}}^{2}\right)$, where $\ell_{\rm B}=e^2/(4\pi \varepsilon_{0}\varepsilon k_{\rm B}T)$ is the Bjerrum length.
When the colloid gets very close to the wall, $d\to0$, the above Derjaguin approximation fails. However, as discussed in Sec.~\ref{Discussion}, this regime does not affect our results.

One can show that the force per unit area, $\Pi(d)$, is independent of $\xi$ and equals to~\cite{Tomer2020, Parsegian1972}
\begin{align} \label{Pressure}
\begin{split}
\Pi(d)& =\frac{\varepsilon\varepsilon_0}{2}\left(-\psi'^{2}\left(\xi;d\right)+\kappa_{\rm D}^{2}\psi^{2}\left(\xi;d\right)\right)\\
& =-\frac{2\sigma_{{\rm c}}\sigma_{{\rm w}}\cosh\left(\kappa_{{\rm D}}d\right)+\sigma_{{\rm c}}^{2}+\sigma_{{\rm w}}^{2}}{2\varepsilon\varepsilon_0\sinh^{2}\left(\kappa_{{\rm D}}d\right)},
\end{split}
\end{align}
where $\psi'=d\psi/d\xi$. Substituting Eq.~(\ref{Pressure}) into Eq.~(\ref{Force}), we obtain the force between the colloid and one surface. Integrating the force, we further get the interaction potential $U_1$ with a single wall, $U_1(d)=-\int_{\delta}^{d} {\rm d} h \, F(h)$ where the lower cutoff $\delta$ is an arbitrary distance.
The total interaction potential of the colloid, when taking into account the two walls is $U(x)=U_1(d_0+x)+U_1(d_0-x)$. After some algebra we obtain,
\begin{align}
\begin{split}
U (x) & =\frac{\pi a}{\varepsilon\varepsilon_0\kappa_{{\rm D}}^{2}}\bigg[2\sigma_{{\rm c}}\sigma_{{\rm w}}\ln\left(\frac{\cosh(\kappa_{{\rm D}}d_{0})+\cosh(\kappa_{{\rm D}}x)}{\cosh(\kappa_{{\rm D}}d_{0})-\cosh(\kappa_{{\rm D}}x)}\right)
\\
 & +\left(\sigma_{{\rm c}}^{2}+\sigma_{{\rm w}}^{2}\right)\ln\left(\frac{\cosh(2\kappa_{{\rm D}}d_{0})}{\cosh(2\kappa_{{\rm D}}d_{0})-\cosh(2\kappa_{{\rm D}}x)}\right)\bigg].
 \end{split}
\end{align}

We note that since $U(x)$ is constructed from integrating $\Pi$ twice, the effective spring constant $K=U''|_{x=0}$ [see Eq.~(\ref{harmonic})], can be directly obtained from $\Pi$,
\begin{equation}
K=4\pi a \Pi(d_0).
\end{equation}

\section{Charged walls with fixed surface electrostatic potential} \label{appendixB}
%%%%%%%%%%%%
Assuming that the walls have a fixed surface electrostatic potential, $V$, we repeat the calculation of Appendix~\ref{appendixA} and Sec.~\ref{Equilibrium}.

For the space between the left wall and the colloid, the boundary conditions are $\psi |_{\xi = 0}=V$ and $\rm{d}\psi/\rm{d}\xi|_{\xi=d} = \sigma_{\rm c}/\left(\varepsilon_0\varepsilon\right)$. The solution to Eq.~(\ref{DH}) is
\begin{equation} \label{Eq_System2b_eq1}
\psi\left(\xi\right)=\frac{V\varepsilon_{0}\varepsilon\kappa_{{\rm D}}\cosh\left[\kappa_{{\rm D}}(d-\xi)\right]+\sigma_{{\rm c}}\sinh\left(\kappa_{{\rm D}}\xi\right)}{\varepsilon\varepsilon_0\kappa_{{\rm D}}\cosh\left(\kappa_{{\rm D}}d\right)},
\end{equation}
and following a similar calculation as done in Appendix A, the total interaction potential is
\begin{widetext}
\begin{align} \label{Eq_System2b_eq2}
\begin{split}
U\left(x\right) & =\frac{\pi a}{\varepsilon\varepsilon_0\kappa^2_{{\rm D}}} \bigg[
\left[\left(V\varepsilon_{0}\varepsilon\kappa_{{\rm D}}\right)^{2}-\sigma_{{\rm c}}^{2}\right]
\ln\left(\cosh[\kappa_{{\rm D}}(d_{0}+x)] \cosh[\kappa_{{\rm D}}(d_{0}-x)]\right)
\\
& -4\sigma_{{\rm c}}V\varepsilon\varepsilon_0\kappa_{{\rm D}}
\left[\tan^{-1}\left(\tanh \frac{\kappa_{\rm D}(d_{0}+x)}{2}\right)
+\tan^{-1}\left(\tanh \frac{\kappa_{\rm D}(d_{0}-x)}{2}\right)\right]\bigg].
\end{split}
\end{align}
\end{widetext}
Qualitatively, the behavior of $U$ when varying $V$ is similar to its behavior in Sec.~\ref{Equilibrium}. The stability condition is
\begin{equation} \label{Eq_System2b_eq5}
\begin{split}
&\,\,\,\,\,\,\,\,\,\,\delta >{\rm e}^{-\kappa_{{\rm D}}d_0}\\
&{\rm or}\\
&\,\,\,\,\,\,\,\,\,\,\delta <-{\rm e}^{\kappa_{{\rm D}}d_0},
\end{split}
\end{equation}
where $\delta=V\varepsilon_{0}\varepsilon\kappa_{{\rm D}}/\sigma_{{\rm c}}$. Unlike Sec.~\ref{Equilibrium}, here the stability condition does not coincide with the condition that the walls repel the colloid at the mid-plane. The latter condition is
\begin{equation} \label{Eq_System2b_eq6}
\begin{split}
&\,\,\,\,\,\,\,\,\,\,\delta  >\frac{1}{{\rm e}^{\kappa_{{\rm D}}d_0}+\sqrt{1+{\rm e}^{2\kappa_{{\rm D}}d_0}}}\\ &{\rm or}\\
&\,\,\,\,\,\,\,\,\,\,\delta  <\frac{1}{{\rm e}^{\kappa_{{\rm D}}d_0}-\sqrt{1+{\rm e}^{2\kappa_{{\rm D}}d_0}}}.
\end{split}
\end{equation}

If the stability condition of Eq.~(\ref{Eq_System2b_eq5}) is satisfied, we can approximate the interaction potential around $x=0$, to be $U\approx K x^2/2$, with the effective spring constant
\begin{equation} \label{Eq_System2b_eq7}
K=\left( \frac{2\pi a}{\varepsilon\varepsilon_0} \right)
\frac{2\sigma_{{\rm c}}V\varepsilon\varepsilon_0\kappa_{{\rm D}}\sinh(\kappa_{{\rm D}}d_{0})+\left(V\varepsilon\varepsilon_0\kappa_{{\rm D}}\right)^{2}-\sigma_{{\rm c}}^{2}}{\cosh^{2}(\kappa_{{\rm D}}d_{0})}.
\end{equation}
As shown in Sec.~\ref{general}, the value of $K$ determines the type of colloid diffusion. For a fixed $\sigma_{\rm c}$, $K$ grows as $V$ is increased, and decays to zero for $\kappa_{\rm D} d_0 \gg 1$.

 %%%%%%%%%%%%%%%%%%%%%%%%%%%%%%%%%%

%%%%%%%%
\end{document}